# Deep ICE: A Deep learning approach for MRI Intracranial Cavity Extraction

José V. Manjón[1], Jose E. Romero[1], Roberto Vivo-Hernando[2], Gregorio Rubio-Navarro[3], María De la Iglesia-Vaya[4], Fernando Aparici-Robles[5] and Pierrick Coupé[6]

[1] Instituto de Aplicaciones de las Tecnologías de la Información y de las Comunicaciones Avanzadas (ITACA), Universitat Politècnica de València, Camino de Vera s/n, 46022 Valencia, Spain.

[2] Instituto de Automática e Informática Industrial, Universitat Politècnica de València, Camino de Vera s/n, 46022 Valencia, Spain.

[3] Departamento de matemática aplicada, Universitat Politècnica de València, Camino de Vera s/n, 46022 Valencia, Spain.

[4] Brain Connectivity Laboratory, Joint Unit FISABIO & Prince Felipe Research Centre (CIPF), Valencia, Spain

[5] Área de Imagen Medica. Hospital Universitario y Politécnico La Fe. Valencia, Spain

[6] CNRS, Univ. Bordeaux, Bordeaux INP, LaBRI, UMR5800, PICTURA Research Group, 351, cours de la Libération F-33405 Talence cedex, France.

**Corresponding author:** José V. Manjón. Instituto de Aplicaciones de las Tecnologías de la Información y de las Comunicaciones Avanzadas (ITACA), Universidad Politécnica de Valencia, Camino de Vera s/n, 46022 Valencia, Spain.

Tel.: (+34) 96 387 70 00 Ext. 75275 Fax: (+34) 96 387 90 09.

E-mail address: jmanjon@fis.upv.es (José V. Manjón)

**Abstract**

Automatic methods for measuring normalized regional brain volumes from MRI data are a key tool to help in the objective diagnostic and follow-up of many neurological diseases. To estimate such regional brain volumes, the intracranial cavity volume is commonly used for normalization. In this paper, we present an accurate and efficient approach to automatically segment the intracranial cavity using a volumetric 3D convolutional neural network and a new 3D patch extraction strategy specially adapted to deal with the traditional low number of training cases available in supervised segmentation and the memory limitations of modern GPUs. The proposed method is compared with recent *state-of-the-art* methods and the results show an excellent accuracy and improved performance in terms of computational burden.

# 1. Introduction

Quantitative brain image analysis rely on different types of methods to accurately quantify the state of the brain through anatomical or functional measures. One of these methods is the segmentation of the entire brain region, which aims to assign a label to each of the voxels of the brain area. This operation has received different names in literature such as brain extraction, skull-stripping or intracranial cavity extraction. In each case, the aim is to isolate the brain or intracranial tissues (depending on area definition) from the raw image.

Usually, the term brain extraction or skull-stripping has been used to describe methods that segment the brain images not including the external cerebrospinal fluid (CSF) (but including the ventricular CSF) in their mask. Intracranial cavity extraction normally refers to the identification of all tissues within the skull, including external CSF. The accurate estimation of the intracranial volume has been shown to be important to obtain robust and reliable normalized measurements of brain structures [1].

There is a large number of brain extraction methods, most of them based on the use of T1 weighted MR images due to their excellent tissue contrast. Some of the most well-known methods are: BET (Brain Extraction Tool) from the FSL image processing library [2] ,3dIntracranial [3], Hybrid Watershed algorithm (HWA) [4], ROBEX [5] and Brain Surface Extractor (BSE) [6]. Another well-known approach for intracranial cavity extraction is to perform a full modeling of brain intensities using a parametric model such as done in Statistical Parametric Mapping (SPM) [7], VBM (http:/dbm.neuro.uni-jena.de/vbm) or more recently CAT (http://dbm.neuro.uni-jena.de/cat/software) package.

More recent brain extraction methods are MAPS [8] and BEaST [9]. Both methods rely on the application of a multi-atlas label fusion strategy. MAPS uses multi-template non-linear registrations followed by a voxel-wise label fusion while BEaST uses a single linear registration to a standardized space (MNI152) in combination with the non-local patch-based label fusion method proposed by Coupe et al. [10]. Both techniques perform very well although MAPS has a much larger computational load compared to BEaST [9]. Finally, for intracranial cavity extraction, NICE method was proposed [11] which represents the current *state of the art* for MRI intracranial cavity extraction. This method is a multi-atlas segmentation method similar to BEaST but faster and more accurate due to the use of a non-local dense patch prediction and multithreading acceleration.

Lately, new methods using deep learning techniques have also been proposed. Due to the limited GPU memory, so far, most of these have been based on a patch-wise strategy or 2D frameworks. However, patch-wise or 2D approaches have limited context information which makes them suboptimal. For example, Kleesiek et al. [12] proposed a novel method based on the use an 8-layer patch-based 3D convolutional neural networks (CNN). A different approach was proposed by Salehi et al. [13] using an Autocontext 2D CNN which uses a U-Net [14] architecture in cascade where the prediction of the first network is used as extended input of the second network.

In this paper, we present a volume-based 3D CNN method for intracranial cavity extraction where we deal with the low number of training cases and the memory limitations of the modern GPUs using a strided decimated volume processing and reconstruction which allows to train 3D volumetric networks more effectively. The use of this approach allows to obtain *state-of-the-art* accuracy in a very efficient way (few seconds).

## 2. Methods

### 2.1. Dataset description

In this study, we used a dataset which consists of 50 manually segmented MR volumes, with ages that cover nearly the entire human life-span [11,15]. Details of the dataset are the following:

- **Normal adults dataset**: 30 normal subjects (age range: 24-75 years) randomly selected from the open access IXI dataset (http://www.brain-development.org/). This dataset contains images from nearly 600 healthy subjects from several hospitals in London (UK). Both 1.5 T (7 cases) and 3 T (23 cases) images were included in this dataset. 3T images were acquired on a Philips Intera 3T scanner (TR = 9.6 ms, TE = 4.6 ms, flip angle=8°, slice thickness=1.2 mm, volume size=256×256x150, voxel dimensions = 0.94×0.94×1.2 mm$^3$). 1.5 T images were acquired on a Philips Gyroscan 1.5T scanner (TR = 9.8 ms, TE = 4.6 ms, flip angle=8°, slice thickness = 1.2 mm, volume size=256×256x150, voxel dimensions = 0.94×0.94×1.2 mm$^3$).

- **Alzheimer Disease (AD) dataset**: 10 patients with Alzheimer's disease (age range= 75-80 years, CDR = 1.1±0.4) scanned using a 1.5 T General Electric Signa HDx MRI scanner (General Electric, Milwaukee, WI) randomly selected. This dataset consisted of high resolution T1-weighted sagittal 3D MP-RAGE images (TR=8.6 ms, TE=3.8 ms, TI=1000 ms, flip angle=8°, slice thickness=1.2 mm, matrix size=256×256, voxel dimensions=0.938×0.938×1.2 mm$^3$). These images were downloaded from the brain segmentation testing protocol [16] website (https://sites.google.com/site/brainseg/) while they belong originally to the open access OASIS dataset (http://www.oasis-brains.org/).

- **Pediatric dataset**: 10 infant subjects were also downloaded from the brain segmentation testing protocol [16] website (https://sites.google.com/site/brainseg/). These data are also available at http://www.brain-development.org (this dataset is property of the Imperial College of Science Technology & Medicine and has been used after accepting the license agreement). The selected 10 cases are from the full sample of 32 two-year old infants born prematurely (age = 24.8 ± 2.4 months). Sagittal T1 weighted volumes were acquired from each subject (1.0 T Phillips HPQ scanner, TR=23 ms, TE=6 ms, slice thickness=1.6 mm, matrix size= 256×256, voxel dimensions= 1.04×1.04×1.6 mm$^3$ re-sliced to isotropic 1.04 mm$^3$).

**2.2. Data preprocessing**

To improve the image quality and to set all the cases in a similar geometrical and intensity space, all images in the dataset were preprocessed as follows: 1) Images are denoised using the Spatially Adaptive Non-Local Means (SANLM) filter [17], 2) then images are inhomogeneity corrected using the N4 method [18], 3) the resulting images are affine registered to the Montreal Neurological Institute (MNI152) space using the ANTS software [19] and 4) the images are intensity normalized by subtracting the mean of the brain area and dividing by its standard deviation (the brain area was roughly estimated from the MNI brain probability map estimated as the mean of all the IC masks of the training images in the MNI space). All the methods were run with their default parameters.

The preprocessed images in the MNI space have a size of 181x217x181 voxels with 1 mm$^3$ voxel resolution. All the images were manually segmented using the ITK-SNAP [25] software as described in [14]. Although some of the previously proposed methods work in the native space, we perform the segmentation in the MNI space to reduce the problem complexity given the low number of training cases. An example of the preprocessed images and the corresponding manual label is shown in figure 1. The manually segmented data used to developed the proposed method and to support the findings of this study have not been made available because it belongs the protected software volBrain.

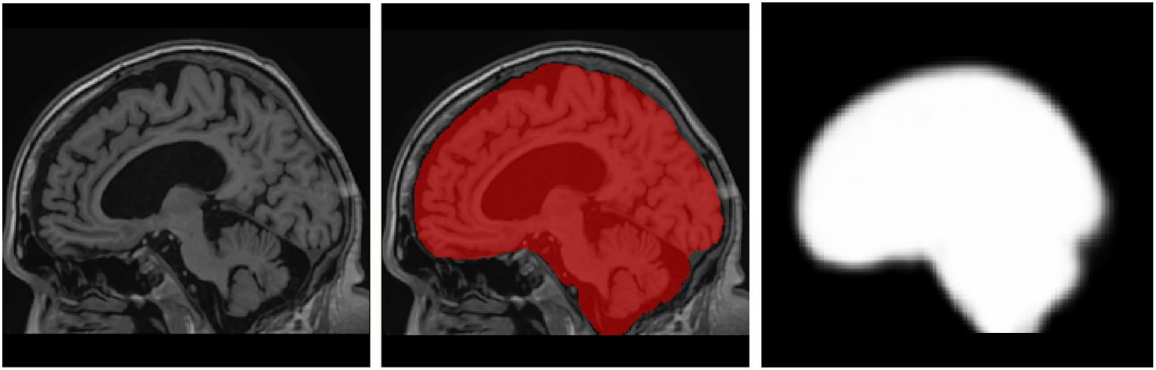

*Figure 1.* Example of one case of the dataset. Left: preprocessed image. Middle: Manual segmentation of the intracranial space (note that all external CSF is included in our mask definition). Right: Intracranial cavity probabilistic a-priori map estimated as the mean of all the IC masks of the training images in the MNI space.

## 2.3. Proposed Method

The proposed method is based on the classical U-Net architecture [14] with some modifications to adapt it to intracranial cavity extraction problem. We used three 3D convolution layers (kernel size of 3x3x3 voxels) per resolution level with a ReLU activation function and batch normalization layers. We used a dropout layer (with rate of 0.5) before each max pooling layer (with factor 2) to minimize overfitting problems. The first resolution level had 48 filters, with following resolution levels increasing on the previous by a factor 2 to compensate for the loss of spatial resolution. Similarly, the number of filters is reduced by 2 in the ascending path at each resolution level. The last block has a convolutional layer has filter size of 1x1x1 and a softmax layer with 2 output channels representing the probabilities of the background and the foreground. In Figure 2 the scheme of the proposed network is shown.

The input of the proposed network consisted of a tensor with two channels. The first channel is the input volume and the second the *a-priori* probability map (computed as the mean IC probability map of the training data). The output is also a tensor of two channels, representing the probabilities of background and foreground. The resulting network had a total of 56 layers and 19,733,286 trainable parameters.

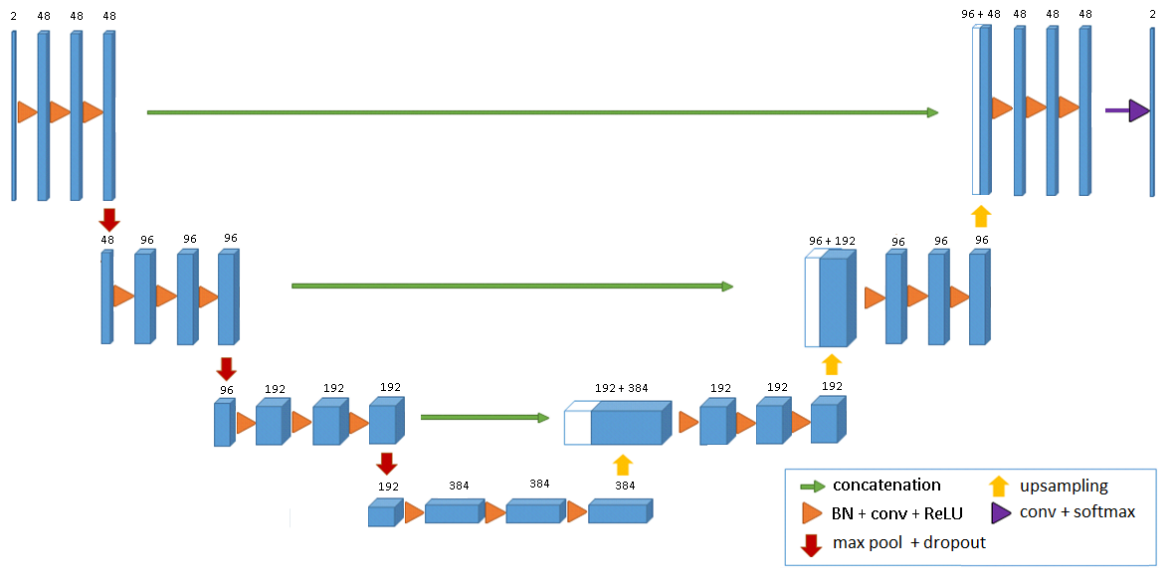

*Figure 2.* Scheme of the proposed modified U-Net CNN.

### *Strided patch extraction and reconstruction*

Training the proposed network with input volumes of 181x217x181 voxels is not possible due to memory limitations of the GPU. One possible solution to this problem is to use patch-wise strategy which sequentially processes different regions of interest by extracting overlapping or non-overlapping patches covering the whole volume of interest. Although this approach is effective, such strategy requires that the network has to learn patterns from patches at different locations. We propose as an alternative to perform a strided patch extraction where we extract also patches but using a stride-based strategy covering the whole volume. In Figure 3, two examples of the proposed strided patch extraction is shown and compared to classical patch extraction (examples are shown in 2D for simplicity but we use 3D extraction in our method). By using a strided extraction, all patches share a similar anatomy (but shifted in the three axes) and therefore the network has to solve a simpler problem compared to the classic patch extraction strategy as can be noted in figure 3.

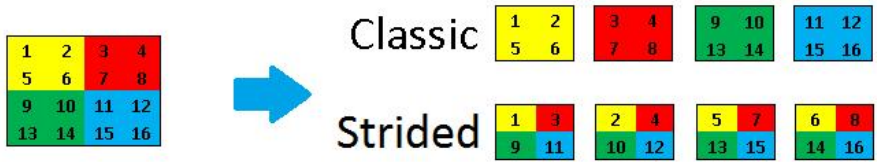

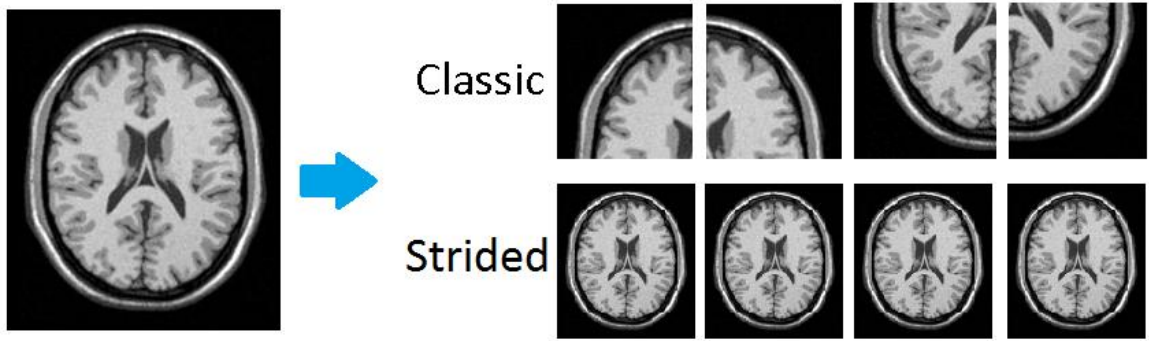

a) Matrix: classic vs strided patch extraction

b) MRI: classic vs strided patch extraction

*Figure 3. a) 2D classic vs strided patch extraction of an example matrix. b) 2D MRI example of classic vs strided patch extraction. In both cases the number of patches is the same (non-overlapping patches) but the information they share is very different.*

Once the network performs its prediction, an inverse strided reconstruction is performed by setting each voxel prediction in its original position. To regularize the reconstructed probability maps (and to correct small reconstruction errors) a small 3D Gaussian filter (sigma=1) is applied.

### *Training*

To train the network we used an *Adam* optimizer (lr=0.001) with a dice loss [20]. A data generator function (with batch size equal to 1) was used to feed the network. Given the low number of training cases, we pre-trained the network using automatic segmentations produced by NICE method [11].

### *Post-training*

We used a callback to save the best network during training using a validation dataset to avoid under and overfitting. We also saved the network weights every two epochs to perform snapshot averaging [21] after the training to further improve the network results.

While averaging several independently trained network predictions has been shown to improve segmentation accuracy this approach is not very efficient since it requires to train several networks and to run several predictions which increase both training and test time. Snapshot averaging enables to improve the network prediction by averaging the network weights of a single training session at different training epochs which provides a much more efficient solution. Differently from Huang *et al.* [21], we included the best network saved in the ensemble averaging.

Finally, at test time, we further improved the results by using an approach recently proposed by Gal and Ghahramani [22] where the dropout layer is kept at test time and several network predictions are averaged for different dropout layer states. This is also called Test Time DropOut (TTDO).

## 3. Experiments

All the experiments were performed using Tensorflow 1.12 and Keras 2.2.4 in a Titan X GPU with 12 GB of memory. Since our dataset consist of only 50 MRI cases, we performed a K-fold cross validation (K=5) to evaluate the proposed method. Each fold was composed by 30 training cases, 10 validation cases and 10 test cases and the network was trained during 20 epochs with 480 steps per epoch. The training and validation cases were processed to extract the strided patches (8 patches per volume) and we perform data augmentation by adding the horizontally mirrored images and segmentations which yield a total of 480 training patches and 160 validation patches per fold. Images were zero padded to have a size of 192x224x192 so the strided patches had a size of 96x112x96 voxels. After segmentation, the resulting segmentation masks were set back to their original position and the final volume was cropped back to the original size of 181x217x181 voxels.

### 3.1. Method settings

To explore the different options of the proposed method, we ran experiments in only one fold (to explore as much options as possible) using the validation data to evaluate the results. No post-training optimizations were performed in these experiments. To compare the segmentation results of the different options we used the DICE coefficient [23] in percentage.

First, we evaluated the proposed network using classic vs strided patch extraction. The same network with the same number of patches was trained during 20 epochs, only the patch extraction was different. The strided option not only achieved a higher accuracy but also had a faster convergence. The mean validation dice after training was 98.89 ± 0.21 for the strided extraction and 98.70 ± 0.36 for classic extraction. This difference was found to be statistically significant ($p<0.05$). In Figure 4, a visual example result of the two options is shown.

We also tested the effect of using the *a-priori* probability map in the input (second channel). In this case, we repeated the same experiment but using only one channel as input (T1 patch). The mean validation dice after training for strided patch extraction was 98.88±0.25 and 98.14±0.58 for classic patch extraction. As can be noticed the *a-priori* information significantly helped the classic patch extraction. This is not surprising as other papers have already remarked the usefulness of the *a-priori* information [13] in segmentation. For the strided patch extraction, the inclusion of the *a-priori* information did not significantly improve the results but made the training faster and the results were more stable (lower variability) and thus, we decided to keep it. We think that the inclusion of *a-priori* information did not improve the results probably due to the coherence of the training patterns shown to the network.

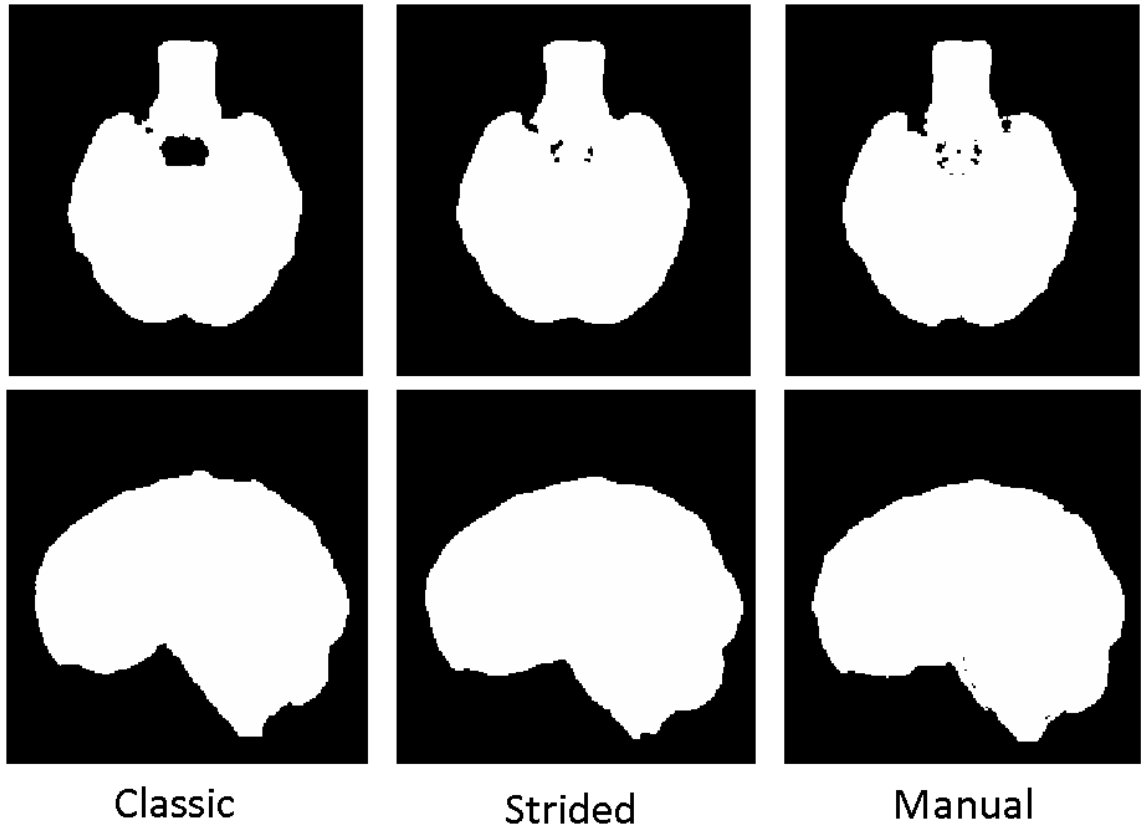

*Figure 4.* Left: Example segmentation result obtained using classic patch extraction strategy. Middle: Example segmentation result using the proposed strided patch extraction. Right: Reference manual segmentation.

## 3.2. Post-training optimization

After training, we can further improve the results using network regularization methods. To show the effect of the snapshot averaging we evaluated the mean validation dice as a function of the number of networks averaged. In Figure 5 (left), we show how the dice improves when we average the best network with the last saved networks (and theoretically more accurate) and how the results degrade when using earlier network versions (less accurate). We saved as the definitive network the one that obtained the best validation results (3 averages in the example).

In a similar way, we evaluated the effect of TTDO as a function of the number of averaged predictions in the validation set. In Figure 5 (right), we show how the dice improves significantly ($p<0.05$) even when using only one prediction (no averaging but with dropout layer active). With 2 predictions results are slightly better but not significantly and later stabilizes. We think that in this case the improved performance of the network can be explained, not from the dropout layer, but from the fact that batch normalization layers behave differently on training and test. To activate dropout layers at test time we set the network on train mode; causing the batch normalization layers to use the current instance mean and variance instead of the saved values during training, which better adapts to the case being processed. Since using more than one prediction did not significantly improve the results we set TTDO with only one prediction for efficiency.

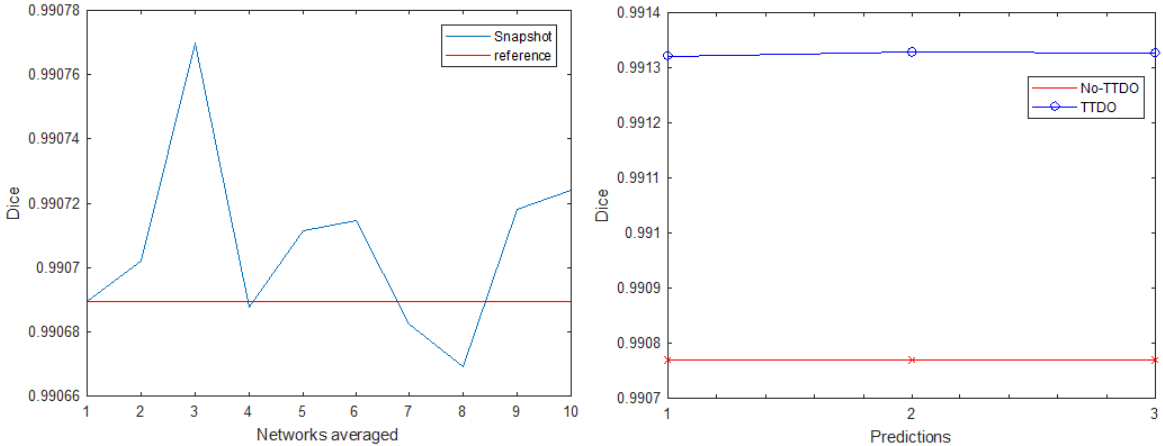

*Figure 5.* Left: Result of snapshot technique. The aggregation of the last networks rapidly improves the dice and later the dice decreases as we include less accurate networks. Right: TTDO results. TTDO provide better results than base prediction but no significant improvement is obtained when using more than one prediction in the averaging.

### 3.3. Final results

To obtain the final results we used the test sets from each of the 5-folds of the dataset. To train the final networks (one per fold), we pre-trained a network using automatic segmentations produced by NICE method using 350 brain MRIs randomly selected from a dataset containing more than 3000 cases used in a recent life-span study [24]. To avoid data contamination, the NICE method used only the training cases of the corresponding fold as manually labeled cases library. We pre-trained the network until convergence using a early stopping criteria (with *patience* parameter set to 5 epochs).

After the pre-training, for each fold, we perform a fine-tuning with the fold training set during 20 epochs. The described snapshot method was used to generate the final network after training. The test time dropout (TTDO) technique was also used to generate the final results. In Table 1, we show the final test results of the network for the different options with and without pre-training to inspect the effect of each part. As can be noted, the pre-training step helps to further improve the results of the proposed method. Although the improvement of snapshot technique was not statistically significant, we decided to keep it since it does not increases the training or testing time and, in general, network aggregation normally helps to improve the generalization capability of the networks.

*Table 1.* *Average DICE coefficient (and standard deviation) for the different training and test options. Second row shows the results without pre-training phase. Best results in bold.*

| Pre-train | + Fine tuning | + Snapshot | + TTDO |
|---|---|---|---|
| 98.58 ±0.44 | 98.96±0.40 | 98.98±0.40 | **99.11± 0.17** |
| N/A | 98.84±0.42 | 98.87±0.41 | 99.04±0.20 |

### 3.4. Method comparison

The proposed method was compared with the state of the art methods BEaST, NICE and VBM8 using the same test dataset. To quantitatively compare the segmentation results we employed the DICE coefficient, sensitivity and specificity measures. Results were also split using the different subsets covering different age ranges to better show the differences in the method's performance. Table 2 summarizes these results. As can be noticed, the overall accuracy of the proposed method is similar to NICE method but with a lower variability (0.20 vs 0.17) which makes it slightly more robust in general. It has also a high sensitivity which means that we could further improve the results by removing some false positives. It is also worth noting that the proposed method performs slightly better than NICE for pediatric cases despite the low number of training cases for this age period.

*Table 2. Average DICE coefficient for the methods compared on the different used datasets. The best results from each column are in bold.*

| Method | Data | All (N=50) | Adults (N=30) | AD (N=10) | Infants (N=10) |
|---|---|---|---|---|---|
| DeepICE | DICE | **99.11±0.17** | 99.20±0.09 | 98.90±0.15 | **99.03±0.14** |
| | SEN | **99.11±0.32** | **99.23±0.21** | 98.86±0.36 | 98.97±0.37 |
| | SPE | 99.69±0.09 | 99.72±0.08 | 99.64±0.15 | **99.68±0.08** |
| NICE | DICE | 99.11±0.20 | **99.21±0.15** | 98.92±0.16 | 98.99±0.19 |
| | SEN | 99.07±0.36 | 99.16±0.35 | **98.87±0.29** | **98.98±0.38** |
| | SPE | **99.71±0.12** | **99.75±0.10** | 99.64±0.14 | 99.65±0.09 |
| BEAST | DICE | 98.80±0.32 | 98.91±0.30 | 98.57±0.18 | 98.66±0.34 |
| | SEN | 98.89±0.62 | 99.02±0.60 | 98.30±0.49 | 99.00±0.50 |
| | SPE | 99.55±0.19 | 99.58±0.17 | 99.60±0.16 | 99.40±0.19 |
| VBM8 | DICE | 97.62±0.52 | 97.88±0.26 | 96.90±0.64 | 97.50±0.33 |
| | SEN | 97.40±0.12 | 97.96±0.51 | 95.87±0.13 | 97.10±0.14 |
| | SPE | 99.26±0.27 | 99.24±0.19 | 99.31±0.33 | 99.26±0.41 |

Finally, execution times of the different methods were compared (excluding preprocessing time). DeepICE took around 3 seconds in a Titan X GPU, the NICE method took around 2 minutes (NICE was implemented as a multithreaded MEX C file), BEaST method took around 25 minutes (we have to note that no multithreading optimizations were used here) and VM8 takes around 8 minutes. NICE, BEaST and VBM8 experiments were performed using MATLAB 2015a 64 bits (Mathworks Inc.) on a desktop PC with an Intel core i7 with 16 GB RAM running windows 7. The full processing time of the proposed DeepICE pipeline (including preprocessing) is around 2 minutes.

## 5. Discussion

We have presented a new deep learning based method for intracranial cavity extraction that we called DeepICE. The proposed method uses 3D strided patches instead of classical patches which allows the method to be trained with a less complex set of patterns making the training more efficient and accurate.

We pre-trained the network using automatic segmentations produced by the NICE method which helped to increase the method's accuracy. In post-training, we used snapshot averaging and test time dropout to further improve the results of the proposed method. While the snapshot technique did not significantly improved the results, TTDO did. However, we found that this improved performance cannot be attributed to the Bayesian prediction resulting from the average of different prediction but from the way that batch normalization layers are working at training and test time when using small batch sizes (N=1 here) since instance means and variances are directly estimated from the input data instead of using historical values saved during training as normally done at test time.

The proposed method was shown to perform better than BEaST and VBM8 methods and performed similar than NICE method (although with consistently lower variability). For age ranges, DeepICE slightly outperformed NICE for pediatric cases and performed slightly worse for AD and normal adult subjects. Although DeepICE method obtained a similar dice to NICE it is much more efficient since it is approximately 40 times faster (around 3 seconds compared to 2 minutes). Another important difference is that the proposed method demonstrated lower variability than NICE.

The proposed method works in MNI space while most of the related deep learning methods work at native space. To work at a standardized space has many benefits. This approach can be seen as a kind of *data "collapse"* approach compared to the classical data augmentation methods where pseudo-random transformations of the images are applied to increase the size of the training dataset. By "collapsing" the data space to a specific resolution and orientation we reduce the complexity of the problem and therefore make the training process more effective.

The accuracy, efficiency and robustness of the proposed method makes it ideal for big data analysis where large amounts of MRI data need to be automatically analyzed. We plan to include the proposed method within our volBrain pipeline [15] to further reduce the processing times of our online service.

**Acknowledgments**

This research was supported by the Spanish DPI2017-87743-R grant from the Ministerio de Economia, Industria y Competitividad of Spain. This work also benefited from the support of the project DeepvolBrain of the French National Research Agency (ANR-18-CE45-0013). This study was achieved within the context of the Laboratory of Excellence TRAIL ANR-10-LABX-57 for the BigDataBrain project. Moreover, we thank the Investments for the future Program IdEx Bordeaux (ANR-10- IDEX- 03- 02, HL-MRI Project), Cluster of excellence CPU and the CNRS. The authors gratefully acknowledge the support of NVIDIA Corporation with their donation of the TITAN X GPU used in this research.